\documentclass[aip,graphicx]{revtex4-2}
\usepackage{graphicx}% Include figure files
\usepackage{dcolumn}% Align table columns on decimal point
\usepackage{bm}% bold math
\usepackage{hyperref}
\usepackage{upgreek}

\begin{document}

\title{Study of the energy spectrum of alpha particles in an experiment \\ on irradiation of a boron target with a proton beam at the Prometheus accelerator} %Title of paper

\author{A. S. Rusetskii }
\email[]{ruseckijas@lebedev.ru}
%\thanks{}
\affiliation{P. N. Lebedev Physical Institute, Leninsky pr. 53, 119991 Moscow, Russia}

\author{M. A. Negodaev}
%\thanks{}
\affiliation{P. N. Lebedev Physical Institute, Leninsky pr. 53, 119991 Moscow, Russia}

\author{A. V. Oginov}
%\email[]{oginov@lebedev.ru}
%\thanks{}
%\altaffiliation{P. N. Lebedev Physical Institute, Leninsky pr. 53, 119991 Moscow, Russia}
%\affiliation{P. N. Lebedev Physical Institute, Leninsky pr. 53, 119991 Moscow, Russia}
\affiliation{P. N. Lebedev Physical Institute, Leninsky pr. 53, 119991 Moscow, Russia}

\author{V. A. Ryabov}
%\thanks{}
%\altaffiliation{P. N. Lebedev Physical Institute, Leninsky pr. 53, 119991 Moscow, Russia}
%\affiliation{P. N. Lebedev Physical Institute, Leninsky pr. 53, 119991 Moscow, Russia}
\affiliation{P. N. Lebedev Physical Institute, Leninsky pr. 53, 119991 Moscow, Russia}

\author{K. V. Shpakov}
%\email[]{}
%\thanks{}
%\altaffiliation{P. N. Lebedev Physical Institute, Leninsky pr. 53, 119991 Moscow, Russia}
%\affiliation{P. N. Lebedev Physical Institute, Leninsky pr. 53, 119991 Moscow, Russia}
\affiliation{P. N. Lebedev Physical Institute, Leninsky pr. 53, 119991 Moscow, Russia}

\author{A. E. Shemyakov}
%\email[]{}
%\thanks{}
%\altaffiliation{P. N. Lebedev Physical Institute, Leninsky pr. 53, 119991 Moscow, Russia}
%\affiliation{P. N. Lebedev Physical Institute, Leninsky pr. 53, 119991 Moscow, Russia}
\affiliation{P. N. Lebedev Physical Institute, Leninsky pr. 53, 119991 Moscow, Russia}

\author{I. N. Zavestovskaya}
%\email[]{}
%\thanks{}
%\altaffiliation{P. N. Lebedev Physical Institute, Leninsky pr. 53, 119991 Moscow, Russia}
%\affiliation{P. N. Lebedev Physical Institute, Leninsky pr. 53, 119991 Moscow, Russia}
\affiliation{P. N. Lebedev Physical Institute, Leninsky pr. 53, 119991 Moscow, Russia}


%\date{\today}
\date{10.11.2024}

\begin{abstract}
The energy spectrum of alpha particles from the nuclear reaction p + $^{11}$B $\rightarrow  3\alpha$ was studied using the beam of the injector of the proton synchrotron of the Prometheus proton therapy complex. The reaction products (alpha particles) were recorded using a CR-39 track detector. The detectors were calibrated using a $^{241}$Am radioactive source. It was determined that mainly alpha particles emitted from the boron target have energies from 3 to 5.5~MeV. In this case, a significant part of the alpha particles emitted from the depths of the target have significantly lower energies compared to the calculated ones due to ionization losses. Measuring the energy spectrum of alpha particles from targets containing boron is of great scientific and practical interest for identifying the mechanisms of boron-proton capture therapy and determining the additional contribution to the therapeutic effect of proton irradiation.
 \end{abstract}


\maketitle %\maketitle must follow title, authors, abstract and \pacs

%\section{}
%\label{}
%\subsection{}
%\subsubsection{}

\section{Introduction}

In research related to the development of proton therapy for oncological diseases, the promising method of boron-proton capture therapy (BPCT), which increases the biological effectiveness of protons, is currently being studied~\cite{1-3}. BPCT is based on the nuclear reaction of boron-proton fusion, discovered in the 1930s by Oliphant and Rutherford~\cite{4}:

\begin{equation}
{\rm p} + {\rm ^{11}B} \rightarrow  3\alpha + 8.7~{\rm MeV}.
\label{eq1}
\end{equation}

As a result of this reaction, a compound nucleus $^{12}$C$^{*}$ is formed in a very short period of time, which is in an excited state and decays into an alpha particle and a beryllium nucleus $^8$Be, which then decays into two alpha particles. This process is exothermic, releasing a total energy of 8.7~MeV in the form of kinetic energy transferred to three alpha particles. Alpha particles generated by the interaction of decelerating protons with $^{11}$B atoms have high radiobiological efficiency directly in tumor tissue, which leads to a more effective therapeutic effect on the affected tissue compared to beam protons. It should be noted that the boron-proton fusion reaction can be considered as an important, but not the only mechanism, since other biological reactions caused by such particles can also play an important role~\cite{2}. At proton energies of 0.1–-5~MeV, the reaction cross section becomes significant, thereby greatly increasing alpha particle production around the Bragg peak region. Of particular interest is the resonance at 675~keV: it has the largest cross section -- 1.4~barn~\cite{5}. This circumstance is one of the advantages of the potential application of reaction (\ref{eq1}) in proton therapy. Thus, measuring the spectrum of alpha particles near the resonance energy emitted from targets containing boron is of great scientific and practical interest for identifying the mechanisms of BPCT, and determining the additional contribution to the therapeutic effect of proton irradiation.

\section{Experimental technique}

\begin{figure*}[ht]
\includegraphics[width=0.45\textwidth]{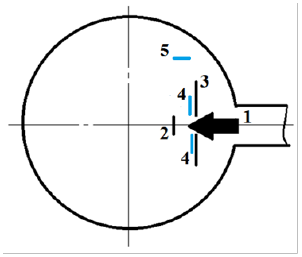}
\vspace{-4mm}
\caption{ \label{fig:1} 
Layout of detectors and targets in a vacuum chamber. 1 -- proton beam, 2 -- target, 3 -- diaphragm (diameter -- 8~mm), 4 and 5 -- CR-39 track detectors.
}
\end{figure*}

The experiment was carried out on the basis of the Prometheus proton accelerator~\cite{6}. An injector was used as a source of protons, which produces a beam with energy of 0.6--1.1~MeV. The experimental scheme is shown in Fig.~\ref{fig:1}.
A beam of protons from the injector with an energy of 0.7~MeV entered the vacuum chamber and hit the target through a diaphragm with a diameter of 8~mm. The initial energy of protons is selected so that the main part of nuclear reactions occurs in a thin surface layer of the target and most alpha particles can escape from it. A plate of natural boron (containing the $^{11}$B isotope - 80\%) was used as a target.
The distances from the center of the target to the middle of track detectors 4 (in Fig.~\ref{fig:1}) $\approx 2$~cm and 5 (in Fig.~\ref{fig:1}) $\approx 10$~cm. Detectors 4 were located on the diaphragm plane parallel to the target surface and particles fell on them at an angle. One of these detectors was coated with 4~$\upmu$m Al to reduce of the loading. The plane of detector 5 (10~cm from the target) was turned towards the target and the particles fell onto it normally. The back side of the detector, located at a distance of 10~cm from the target, was turned towards the wall of the vacuum chamber. The front and back sides of the detector are equally capable of recording the tracks of protons and alpha particles.

Charged particles (protons and alpha particles) were recorded using CR-39 track detectors. The detectors were calibrated using a $^{241}$Am radioactive source in the experimental geometry, when the source is placed in a vacuum chamber in place of the target. In another version, calibration detectors were located in air at various distances from the source and recorded alpha particles with energy losses.

After irradiation, the track detectors were etched in a 6M NaOH solution at 70$^\circ$C for 7~hours. After etching, the detectors were examined under a Nikon L200 microscope with a 50x objective. Lighting was in Dark-field mode. Light passes through the lens, strikes a transparent detector, passes through it, reflects off white matte paper glued to the surface of the slide, and illuminates the surface of the detector from below. The result is circular lateral diffused lighting, in which all surface irregularities are shaded.

To determine the energy spectrum of alpha particles emitted from a boron target, tracks with the most circular shape were selected and compared with calibration tracks.

In one bunch, $\approx 1.5 \cdot 10^{10}$~protons hit the target. Proton energy $\approx 0.7$~MeV. Exposure time is 3~minutes, reset occurs once every 2~seconds. In just 3~minutes, N$_{p} \approx 1.35 \cdot 10^{12}$ protons hit the target.

Two irradiations were carried out -- with a boron target (3~min) and with an aluminum target (3~min). Irradiation with an aluminum target was carried out to measure the yield of beam protons scattered from the target and its contribution to the signal from alpha particles.

The microscope stores RGB (red-green-blue) color images. Tracks are analyzed using the Analyze Particles tool, which finds all tracks, determines their area, determines two roundness indicators: Circularity = $4\pi S/P^{2}$, Roundness = $4S/(\pi R^2)$, where $S$ is area, $P$ is perimeter, $R$ is maximum radius. The parameters of all found areas are written to a {\verb csv} file. For visual control, the contours of all found areas are plotted on the original image and saved separately.

The tracks of alpha particles and protons vary in size. Based on the nature of the calibration results, area distribution, and also from the fact that predominantly proton tracks were detected on detectors irradiated with an aluminum target, it was found that proton tracks have characteristic areas:
from 0.7~$\upmu m^2$ to 5.9~$\upmu$m$^2$, and alpha particles -- from 6.0~$\upmu$m$^2$ to 30.0~$\upmu$m$^2$.

In this experimental setup, it was necessary to take into account the contribution to the overall statistics from defects on the detector surface (with dimensions comparable to real tracks) and the contribution from protons that reach the detectors after scattering from the walls of the vacuum chamber. For this, we used readings from the back sides of the detectors turned towards the chamber wall, which were located at a distance of 10~cm from the target and approximately the same distance from the wall.

From the readings of detectors looking at the chamber walls, the contribution of surface defects and protons scattered from the chamber walls to the overall track statistics was obtained: protons (background) -- $5.03 \cdot 10^5$~cm$^{-2}$, alpha particles (background) -- $4.1 \cdot 10^4$~cm$^{-2}$.

\section{Experimental results and discussion}

\begin{figure*}[ht]
\includegraphics[width=0.55\textwidth]{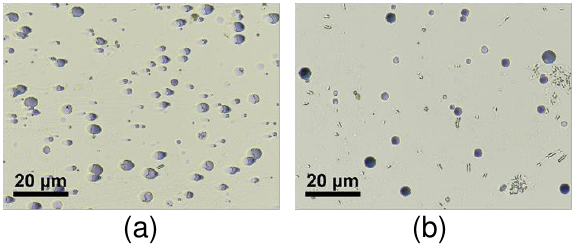}
\vspace{-4mm}
\caption{ \label{fig:2} 
Photomicrographs of the surface of track detectors 7 (a) and 9 (b).
}
\end{figure*}

\begin{figure*}[ht]
\includegraphics[width=0.8\textwidth]{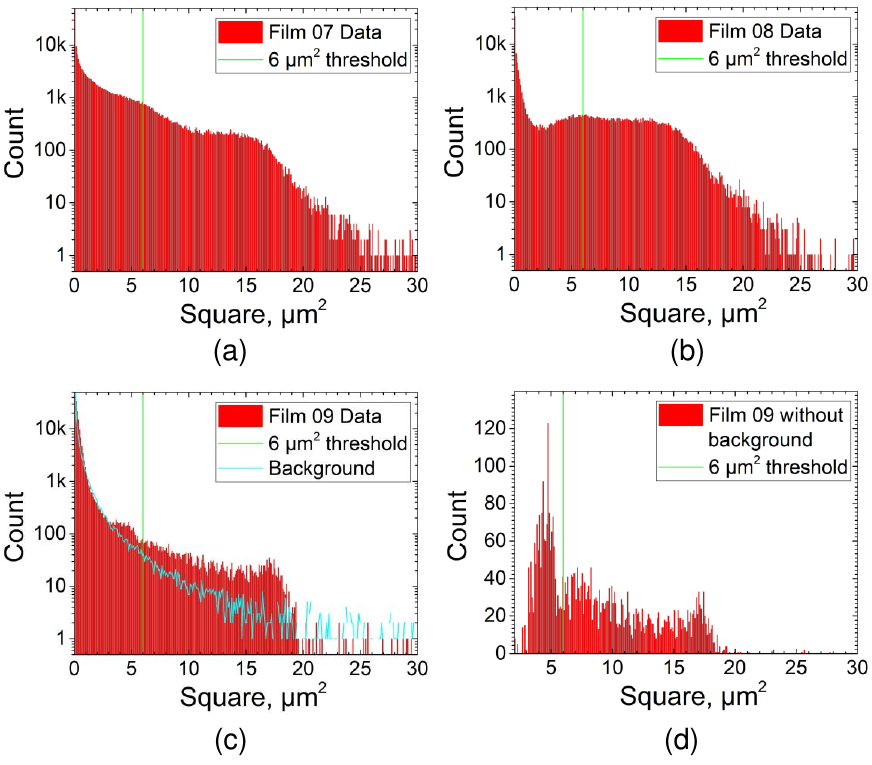}
\vspace{-4mm}
\caption{ \label{fig:3} 
Distributions of track areas for detectors 7 (a), 8 (b) and 9 (c). Difference of foreground and background counts for detector 9.
}
\end{figure*}

Detectors irradiated with an aluminum target are background to detectors irradiated with a boron target. In addition, the background for each detector is the contribution from defects and beam protons scattered from the chamber walls. Figure~\ref{fig:2} shows photomicrographs of the surface of track detectors 7 (left) and 9 (right). Figure~\ref{fig:3} shows histograms of the resulting areas of surface objects (tracks and various defects) for detectors 7, 8 and 9. The average flux of protons and alpha particles is obtained by subtracting the average “background” from the processing results of each specific detector. It can be seen that the main part of the alpha particle tracks have areas in the range from 6 to 20 $\upmu$m2. For detectors 7 and 8, there is no clearly defined peak in the spectrum, since particles fell on them at different angles and the significant contribution of oval tracks distorts the energy spectrum.

\begin{figure*}[ht]
\includegraphics[width=0.85\textwidth]{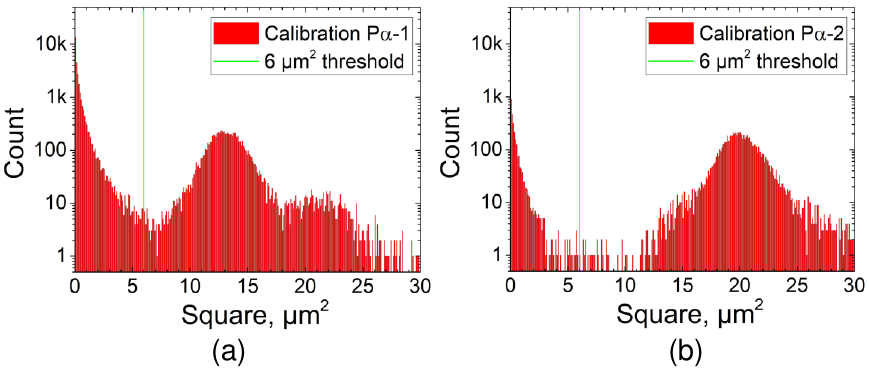}
\vspace{-4mm}
\caption{ \label{fig:4} 
Distributions of the areas of circular tracks for the P$_{\alpha}$-1 (a) and P$_{\alpha}$-2 (b) detectors irradiated with alpha particles from the $^{241}$Am source. (a) Irradiation in vacuum, source-detector distance -1.5~cm. (b) Irradiation in air, source-detector distance - 2.5~cm.
}
\end{figure*}

On detector 9, located 10~cm from the target, particles fell mainly along the normal and the percentage of round tracks was maximum. The distribution for detector 9 (Fig.~\ref{fig:3}c,d) shows a fuzzy peak in the range of 10--20~$\upmu$m$^2$ (average value $\approx 17$ $\upmu$m$^2$).
Figure~\ref{fig:4} shows the distributions of the areas of circular tracks for the P$_{\alpha}$-1 (a) and P$_{\alpha}$-2 (b) detectors irradiated with alpha particles from the $^{241}$Am source. For detector~9 irradiated in vacuum (source-detector distance -- 1.5~cm), a peak is observed at 12.6~$\upmu$m$^2$, which corresponds to the energy of alpha particles with an energy of 5.5~MeV. For a detector irradiated in air, at a source-detector distance of 2.5~cm, a peak is observed at 20~$\upmu$m$^2$, corresponding to the energy of alpha particles with an energy of 3~MeV.

Comparison of the distribution of track areas for detector 9 (Fig.~\ref{fig:2} c) with the distributions of calibration detectors (Fig.~\ref{fig:3} a, b) allows us to conclude that the main part of the alpha particles emitted from the boron target had energies in the range of 3.0--5.5~MeV.

\begin{figure*}[ht]
\includegraphics[width=0.5\textwidth]{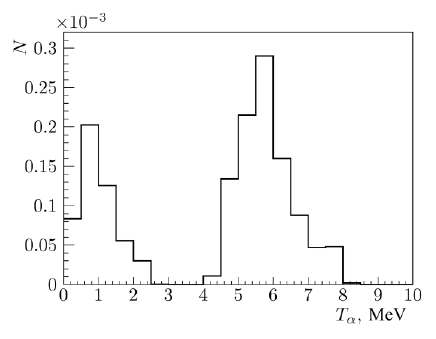}
\vspace{-4mm}
\caption{ \label{fig:5} 
Energy spectrum of final alpha particles obtained by modeling the p + $^{11}$B $\rightarrow  3\alpha$ reaction in~\cite{7}.
}
\end{figure*}

Figure~\ref{fig:5} shows the energy spectrum of final alpha particles obtained by simulating the p + $^{11}$B $\rightarrow  3\alpha$ reaction in~\cite{7}. Since the maximum reaction cross section occurs at proton energies less than 1~MeV, the maximum near 1~MeV corresponds to the first alpha particle of the disruption. The second maximum at about 6~MeV corresponds to alpha particles from the decay of $^8$Be with the release of a binding energy of 8.7~MeV.

A comparison of the rough energy spectrum of alpha particles emitted from a boron target in our experiment with the simulation results shows that we detected particles with energies significantly lower than in Fig.~\ref{fig:5}. This is explained by the fact that alpha particles were born deep in the boron target and reached the detector with significant energy losses. Also, the fact that alpha particles were born at different depths of the target leads to a large spread of their energy when recorded by the detector.

A study of the yield of the nuclear reaction p + $^{11}$B $\rightarrow  3\alpha$ near the resonance energy using CR-39 track detectors was carried out in [8]. The yield of alpha particles from a boron target is estimated at $\approx 10^{-4}$ per proton.

This allows us to conclude that the contribution of alpha particles from reaction (\ref{eq1}) to the total radiation dose is not significant. The enhanced therapeutic effect from the presence of boron is probably explained by other (possibly non-radiation) effects.

\section{Conclusion}

Thus, a study was carried out of the spectrum of alpha particles of the nuclear reaction p + $^{11}$B $\rightarrow  3\alpha$ near the proton resonance energy of 675~keV on the beam of the proton synchrotron injector of the Prometheus proton therapy complex and a natural boron target. The reaction products (alpha particles) were recorded using a CR-39 track detector. The experimentally measured energy spectrum of alpha particles shows that particles with energies in the range of 3.0--5.5~MeV were detected. Comparison with model calculations of the energy spectrum shows that the particles were ejected from the depths of the target, and this led to significant energy losses and ``blurring'' of the spectrum. To determine the overall therapeutic effectiveness of BPCT and the contribution of the boron-proton capture reaction to it, additional invitro\&invivo experiments are required to identify the accompanying physicochemical and radiobiological mechanisms occurring in cancer cells.

%\section*{Supplementary Material}
%See supplementary material for the more detail description of both the mechanism of virtual cathode formation and specifics of hard X-ray release from inter-electrode cluster ensembles.

%\begin{acknowledgments}
%The experimental work of the study by A.V.~Oginov was partially supported by the Russian Scientific Foundation  (grant no.~23-19-00524).
%\end{acknowledgments}

\section*{Data Availability}
The data that supports the findings of this study are available within the article.

%\bibliographystyle{ieeetr}

%\bibliography{kurilenkov_psst}

\end{document}